\documentclass[twocolumn,english]{revtex4-1}
\usepackage[latin9]{inputenc}
\usepackage{xcolor}
\usepackage{babel}
\usepackage{mathtools}
\usepackage{amstext}
\usepackage{amssymb}
\usepackage{graphicx}
\usepackage{esint}
\PassOptionsToPackage{normalem}{ulem}
\usepackage{ulem}
\usepackage[unicode=true,pdfusetitle,
 bookmarks=true,bookmarksnumbered=false,bookmarksopen=false,
 breaklinks=false,pdfborder={0 0 1},backref=false,colorlinks=true]
 {hyperref}
\hypersetup{
 citecolor=blue}
\usepackage{breakurl}

\makeatletter

\providecolor{lyxadded}{rgb}{0,0,1}
\providecolor{lyxdeleted}{rgb}{1,0,0}
\usepackage{chngcntr}
\counterwithout{figure}{section}
\counterwithout{equation}{section}
\usepackage{times}%makes the font bolder
\usepackage{graphicx}
\usepackage{float}
\usepackage{stmaryrd}

\makeatother

\begin{document}
\global\long\def\jm{J_{m}}
\global\long\def\dg{\mathbf{F}}
 \global\long\def\dgcomp#1{F_{#1}}
 \global\long\def\piola{\mathbf{P}}
 \global\long\def\refbody{\Omega_{0}}
 \global\long\def\refbnd{\partial\refbody}
 \global\long\def\bnd{\partial\Omega}
 \global\long\def\rcg{\mathbf{C}}
 \global\long\def\lcg{\mathbf{b}}
 \global\long\def\rcgcomp#1{C_{#1}}
 \global\long\def\cronck#1{\delta_{#1}}
 \global\long\def\lcgcomp#1{b_{#1}}
 \global\long\def\deformation{\boldsymbol{\chi}}
 \global\long\def\dgt{\dg^{\mathrm{T}}}
 \global\long\def\idgcomp#1{F_{#1}^{-1}}
 \global\long\def\velocity{\mathbf{v}}
 \global\long\def\accel{\mathbf{a}}
 \global\long\def\vg{\mathbf{l}}
 \global\long\def\idg{\dg^{-1}}
 \global\long\def\cauchycomp#1{\sigma_{#1}}
 \global\long\def\idgt{\dg^{\mathrm{-T}}}
 \global\long\def\cauchy{\boldsymbol{\sigma}}
 \global\long\def\normal{\mathbf{n}}
 \global\long\def\normall{\mathbf{N}}
 \global\long\def\traction{\mathbf{t}}
 \global\long\def\tractionl{\mathbf{t}_{L}}
 \global\long\def\ed{\mathbf{d}}
 \global\long\def\edcomp#1{d_{#1}}
 \global\long\def\edl{\mathbf{D}}
 \global\long\def\edlcomp#1{D_{#1}}
 \global\long\def\ef{\mathbf{e}}
 \global\long\def\efcomp#1{e_{#1}}
 \global\long\def\efl{\mathbf{E}}
 \global\long\def\freech{q_{e}}
 \global\long\def\surfacech{w_{e}}
 \global\long\def\outer#1{#1^{\star}}
 \global\long\def\perm{\epsilon_{0}}
 \global\long\def\matper{\epsilon}
 \global\long\def\jump#1{\llbracket#1\rrbracket}
 \global\long\def\identity{\mathbf{I}}
 \global\long\def\area{\mathrm{d}a}
 \global\long\def\areal{\mathrm{d}A}
 \global\long\def\refsys{\mathbf{X}}
 \global\long\def\Grad{\nabla_{\refsys}}
 \global\long\def\grad{\nabla}
 \global\long\def\divg{\nabla\cdot}
 \global\long\def\Div{\nabla_{\refsys}}
 \global\long\def\derivative#1#2{\frac{\partial#1}{\partial#2}}
 \global\long\def\aef{\Psi}
 \global\long\def\dltendl{\edl\otimes\edl}
 \global\long\def\tr#1{\mathrm{tr}#1}
 \global\long\def\ii#1{I_{#1}}
 \global\long\def\dh{\hat{D}}
 \global\long\def\inc#1{\dot{#1}}
 \global\long\def\sys{\mathbf{x}}
 \global\long\def\curl{\nabla}
 \global\long\def\Curl{\nabla_{\refsys}}
 \global\long\def\piolaincpush{\boldsymbol{\Sigma}}
 \global\long\def\piolaincpushcomp#1{\Sigma_{#1}}
 \global\long\def\edlincpush{\check{\mathbf{d}}}
 \global\long\def\edlincpushcomp#1{\check{d}_{#1}}
 \global\long\def\efincpush{\check{\mathbf{e}}}
 \global\long\def\efincpushcomp#1{\check{e}_{#1}}
 \global\long\def\elaspush{\boldsymbol{\mathcal{C}}}
 \global\long\def\elecpush{\boldsymbol{\mathcal{A}}}
 \global\long\def\elaselecpush{\boldsymbol{\mathcal{B}}}
 \global\long\def\disgrad{\mathbf{h}}
 \global\long\def\disgradcomp#1{h_{#1}}
 \global\long\def\trans#1{#1^{\mathrm{T}}}
 \global\long\def\phase#1{#1^{\left(p\right)}}
 \global\long\def\elecpushcomp#1{\mathcal{A}_{#1}}
 \global\long\def\elaselecpushcomp#1{\mathcal{B}_{#1}}
 \global\long\def\elaspushcomp#1{\mathcal{C}_{#1}}
 \global\long\def\dnh{\aef_{DH}}
 \global\long\def\dnhc{\mu\lambda^{2}}
 \global\long\def\dnhcc{\frac{\mu}{\lambda^{2}}+\frac{1}{\matper}d_{2}^{2}}
 \global\long\def\dnhb{\frac{1}{\matper}d_{2}}
 \global\long\def\afreq{\omega}
 \global\long\def\dispot{\phi}
 \global\long\def\edpot{\varphi}
 \global\long\def\afreqh{\hat{\afreq}}
 \global\long\def\phasespeed{c}
 \global\long\def\bulkspeed{c_{B}}
 \global\long\def\speedh{\hat{c}}
 \global\long\def\dhth{\dh_{th}}
 \global\long\def\bulkspeedl{\bulkspeed_{\lambda}}
 \global\long\def\khth{\hat{k}_{th}}
 \global\long\def\p#1{#1^{\left(p\right)}}
 \global\long\def\maxinccomp#1{\inc{\outer{\sigma}}_{#1}}
 \global\long\def\maxcomp#1{\outer{\sigma}_{#1}}
 \global\long\def\relper{\matper_{r}}
 \global\long\def\sdh{\hat{d}}
 \global\long\def\iee{\varphi}
 \global\long\def\effectivemu{\tilde{\mu}}
 \global\long\def\fb#1{#1^{\left(1\right)}}
 \global\long\def\mt#1{#1^{\left(2\right)}}
 \global\long\def\phs#1{#1^{\left(p\right)}}
 \global\long\def\thc{h}
 \global\long\def\state{\mathbf{s}}
 \global\long\def\harmonicper{\breve{\matper}}
 \global\long\def\kb{k_{B}}
 \global\long\def\cb{\bar{c}}
 \global\long\def\mb{\bar{\mu}}
 \global\long\def\rb{\bar{\rho}}
 \global\long\def\wavenumber{k}
 \global\long\def\rh{\hat{\mathbf{r}}}
 \global\long\def\zh{\hat{\mathbf{z}}}
 \global\long\def\th{\hat{\mathbf{\theta}}}
 \global\long\def\lz{\lambda_{z}}
 \global\long\def\lt{\lambda_{\theta}}
 \global\long\def\lr{\lambda_{r}}
 \global\long\def\st{\Omega}
 \global\long\def\stz{\Psi}
 \global\long\def\ste{\varphi}
 \global\long\def\stze{\phi}
 \global\long\def\lap{\mathcal{M}}
 \global\long\def\vh{\hat{V}}
 \global\long\def\ch{\hat{c}}
 \global\long\def\wh{\hat{\omega}}
 \global\long\def\rb{\bar{r}}
 \global\long\def\cthick{h}
 \global\long\def\vth{\Delta\vh_{th}}
 \global\long\def\kco{\kh_{co}}
 \global\long\def\normv{\Delta\hat{V}}
\global\long\def\qh{\hat{q}_{A}}
\global\long\def\kh{\hat{k}}
\global\long\def\lzt{\tilde{\lambda}_{z}}
\global\long\def\cratio{\gamma}
\global\long\def\torusvar{\zeta}
\global\long\def\torusfun{\mathcal{T}}
\global\long\def\gapdensity{P}
\global\long\def\gapdom{\mathbb{D}}
\global\long\def\torus{\mathbb{T}}
\global\long\def\se{\Psi}
\global\long\def\slop{a}

\title{Universality and optimality of band-gaps in laminated media}

\author{Gal Shmuel$^{1}$ and Ram Band$^{2}$}

\affiliation{$^{1}$Faculty of Mechanical Engineering, Technion--Israel Institute
of Technology, Haifa 32000, Israel\\
$^{2}$Department of Mathematics, Technion--Israel Institute of Technology,
Haifa 32000, Israel}
\begin{abstract}
We find that the frequency spectra of layered phononic and photonic
composites admit a universal structure, independent of the geometry
of the periodic-cell and the specific physical properties. We show
that this representation extends to highly deformable and multi-physical
materials of tunable spectra. The structure is employed to establish
universal properties of the corresponding band-gaps, and to rigorously
determine their statistical and optimal characteristics. Specifically,
we investigate the density of the gaps, their maximal width and expected
value. As a result, rules for tailoring the laminate according to
desired spectra properties follow. Our representation further facilitates
characterizing the tunability of the band-structure of soft and multi-physical
materials.
\end{abstract}
\maketitle
Wave propagation in heterogeneous media has fascinated the scientific
community for decades. The inhomogeneity causes multiple scattering,
and, in turn, wave interferences that give rise to intriguing phenomena
in various fields. Of particular interest are the transition of conducting
to isolating behavior of electronic crystals \citep{Anderson1958},
localization of electromagnetic waves in dielectrics \citep{yab93},
and attenuation of mechanical motions \citep{Kushwaha1993,Sigalas1995,garcia00prl,vasseur01prlbg,Vasseur08,painter09}
in elastic media. The significance of the latter stems from its central
role in numerous applications; transducers \citep{smith91auld}, waveguides
\citep{miyashita05waveguide}, vibration filters \citep{Khelif03},
acoustic imaging for medical ultrasound and nondestructive testing
\citep{olsson09phononic}, noise reduction \citep{elser06prl} and
cloaking \citep{milton06cloaking,Colquitt2014} are just a few examples.
The mathematical and physical richness of elastic waves in heterogeneous
materials emanates from their vectorial nature, and their spatial
dependency on additional constituents parameters. 

Layered media have been extensively studied \citep{adams08,Gomopoulos10nanoletters,psarobas10prb,walker10apl,Schneider12nano,Schneider13prl,rudykh14prl},
in virtue of their relative simplicity of fabrication and theoretical
modeling. This letter provides new insights on the relation between
their geometry, physical properties, and frequency spectrum. We find
a universal representation for the spectrum, \emph{independent} of
the unit-cell geometry and specific constituents properties. This
structure reveals a universal property of the gaps-density, namely,
its invariance under the change of various geometric and physical
properties. We utilize this representation to determine exactly the
density of the gaps, and their expected and maximal widths. These
results are identified with classes of compositions, hence provide
rules for tailoring the laminate according to desired spectra characteristics.
Thus far, such calculations would necessitate the truncation of infinite
spectra, thereby leading to estimates rather than accurate results. 

The canvas upon which the analysis is presented is of phononic crystals.
The conclusions we draw, however, extend to additional systems. In
virtue of the similarity between electromagnetic and elastodynamics
wave equations for the considered geometry, our insights apply to
one-dimensional (1D) photonic crystals as well. Stratified piezoelectrics
admit a similar spectrum too \citep{Qian04_eq}. Our analysis further
applies to soft non-linear media and multi-physical composites, whose
physical properties are changed upon application of external stimuli.
By these means, the frequency spectrum of such materials is rendered
tunable. Our approach facilitates characterizing this tunability,
as demonstrated in the forthcoming. 

\begin{figure}[t]
\includegraphics[scale=0.35]{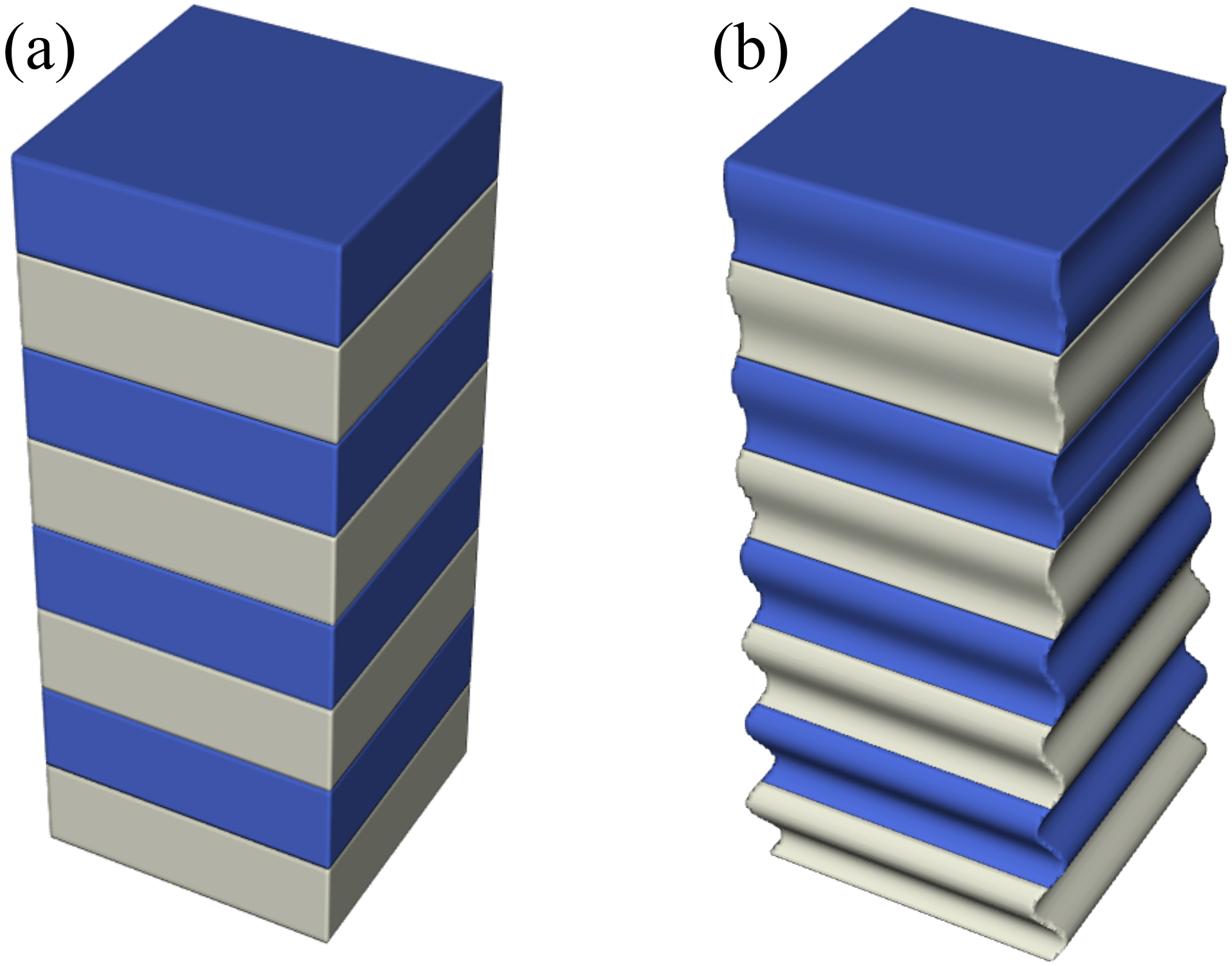}\protect\caption{(a) A 1D crystal of alternated layers. (b) Attenuating waves, associated
with frequencies for which $\left|\eta\right|>1$.}

{\small{}\label{laminate}} 
\end{figure}
We begin by considering a 1D crystal made out of two alternating phases
(Fig. \ref{laminate}a). We denote the phases with $1$ and $2$,
and their associated quantities with superscript $\left(p\right)$,
$p=1$ and $2$, respectively. The dispersion relation governing the
propagation of waves in the crystal is \citep{Yeh77,adams08,Schneider12nano}
\begin{equation}
\arccos\eta=\kb h,\label{eq:dispersion}
\end{equation}
where
\begin{equation}
\eta=\cos\frac{\omega\fb h}{\fb c}\cos\frac{\omega\mt h}{\mt c}-\cratio\sin\frac{\omega\fb h}{\fb c}\sin\frac{\omega\mt h}{\mt c}.\label{eq:eta of d}
\end{equation}
In the above, $\omega$ is the frequency, $k_{B}$ is the Bloch-parameter\emph{,}
$h^{\left(p\right)}$ is the thickness, and $h=h^{\left(1\right)}+h^{\left(2\right)}$.
The parameter $\phase c$ corresponds to different velocities, depending
on the type of media and waves considered. Specifically, when the
crystal is photonic, $\phase c$ is the velocity of light in the phase;
when the crystal is phononic, $\phase c$ is the velocity of either
transverse waves or longitudinal waves, propagating in a bulk. The
parameter $\cratio=\frac{1}{2}\left(\alpha\fb c/\mt c+\alpha^{-1}\mt c/\fb c\right)$
quantifies the contrast between the constituents impedance, where
$\alpha=1$ in the photonic case, and $\alpha=\fb{\rho}/\mt{\rho}$
in the phononic case, $\phase{\rho}$ being the mass density. The
frequency spectrum is obtained by solving Eq. (\ref{eq:dispersion})
for values of\emph{ }$k_{B}$ in the irreducible $1^{\mathrm{st}}$
Brillouin zone \citep{Farzbod2011}, $0\leq k_{B}h\leq\pi$\emph{.
Band-gaps} correspond to ranges of frequencies of attenuating waves,
for which $\left|\eta\right|>1$ (Fig. \ref{laminate}b). 

\begin{figure}[t]
\includegraphics[scale=0.35]{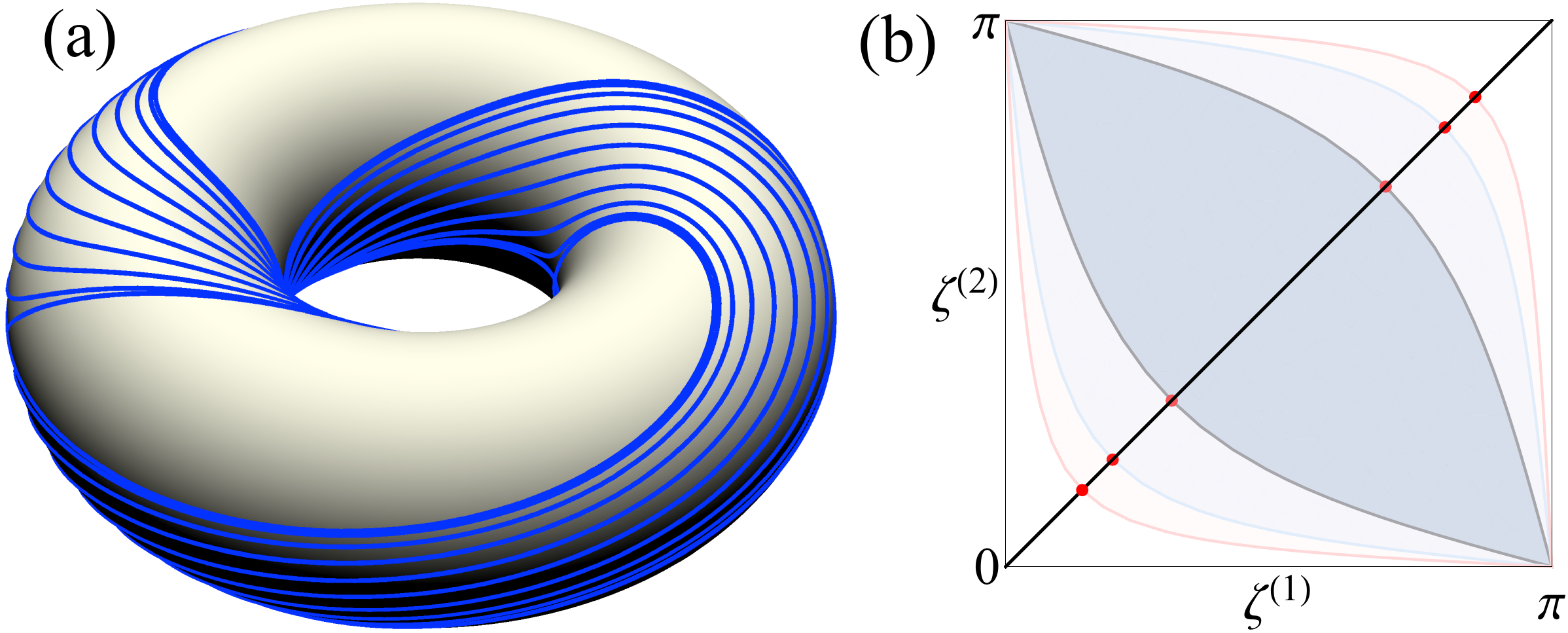}\protect\caption{(color online). (a) Representative contours of the dispersion relation
of real $\protect\kb$, at $\gamma=5$, on the unfolded torus. (b)
Representative domains of $\protect\gapdom$, at exemplary values
of $\gamma=2,5$ and 10, denoted by the gray, blue, and red regions,
respectively, corresponding to ranges of band-gaps. The red dots designate
the intersections of $\protect\gapdom$ with the ray emanating from
the origin, identified with the $1^{\mathrm{st}}$ gap.}

{\small{}\label{torus}} 
\end{figure}

Our analysis of the spectrum benefits from a technique developed for
the study of Schr\"{o}dinger operators on metric graphs \citep{BarGas_jsp00,BerWin_tams10,band13prl}.
The systems we consider necessitate a suitable variation of that approach,
as follows. Upon defining the variables 
\begin{equation}
\phs{\torusvar}\coloneqq\frac{\omega h^{\left(p\right)}}{\phs c},\label{eq:change of variables}
\end{equation}
one can write $\eta$ as a doubly $2\pi$-periodic function of $\fb{\torusvar}$
and $\mt{\zeta}$, namely 
\begin{equation}
\eta\left(\fb{\torusvar},\mt{\torusvar};~\cratio\right)\coloneqq\cos\fb{\torusvar}\cos\mt{\torusvar}-\cratio\sin\fb{\torusvar}\sin\mt{\torusvar}.\label{eq:function for torus}
\end{equation}
Fixing $\gamma$, we consider $\eta$ as a function defined on a 2D
torus of edge length $2\pi$, characterized by the coordinates $\left\{ \fb{\torusvar},\mt{\torusvar}\right\} $.
The absolute value $\left|\eta\right|$ is invariant under the transformations
$\phs{\torusvar}\rightarrow\phs{\torusvar}+\pi$. This symmetry allows
to make a further reduction, and fold the torus into a $\pi$-periodic
torus; we denote the new torus by $\torus$, and denote by $\gapdom$
its subdomain where $\left|\eta\right|>1$ , \emph{i.e.}, where Eq.
(\ref{eq:dispersion}) is solved with imaginary $\kb$. For illustration,
representative contours of the dispersion relation of real $\kb$
are plotted in Fig. \ref{torus}a, on the unfolded torus, at $\gamma=5$.
Also, representative domains of $\gapdom$, at exemplary values of
$\gamma=2,5$ and 10, are depicted in Fig. \ref{torus}b, by the gray,
blue, and red regions, respectively. 

This geometric representation engenders new insights on universal
characteristics of the frequency spectrum, as shown in the sequel.
Towards this end, we interpret Eq. (\ref{eq:change of variables})
as one which defines the following flow on $\torus$
\begin{equation}
\vec{\zeta}\left(\omega\right)=\omega\cdot\left(\frac{\fb h}{\fb c},\frac{\mt h}{\mt c}\right)~~\textrm{mod}~\pi,\label{eq:flow_on_torus}
\end{equation}
where $\omega$ has the role of a time-like parameter. Band-gaps are
identified with values of $\omega$ for which the flow $\vec{\zeta}\left(\omega\right)$
coincides with $\gapdom$. We denote the lower and upper curves bounding
$\gapdom$ by $\mathcal{C}_{l}$ and $\mathcal{C}_{u}$, respectively.
We find that $\eta=-1$ on these curves \footnote{Due to the torus folding, $\eta\leq1$ on $\torus$, and therefore
$\eta<-1$ on $\gapdom$.}; plugging this value into Eq. (\ref{eq:function for torus}) provides
the following expressions for the curves 
\begin{equation}
\mt{\torusvar}=\pi-\arccos\left[\frac{\cos\fb{\torusvar}\pm\gamma\sqrt{\gamma^{2}-1}\sin^{2}\fb{\torusvar}}{1+\left(\gamma^{2}-1\right)\sin^{2}\fb{\torusvar}}\right],\label{eq:bounding curves}
\end{equation}
where the upper (resp. lower) curve $\mathcal{C}_{u}$ (resp. $\mathcal{C}_{l}$),
corresponds to the plus (resp. minus) sign in the numerator. The function
$\eta$ on $\torus$ is invariant under $\pi$-rotation of $\torus$
around its middle, given by the transformation $\left(\fb{\torusvar},\mt{\torusvar}\right)\rightarrow\left(\pi-\fb{\torusvar},\pi-\mt{\torusvar}\right)$,
and under reflection across the line $\mt{\torusvar}=\pi-\fb{\torusvar}$,
by the transformation $\left(\fb{\torusvar},\mt{\torusvar}\right)\rightarrow\left(\pi-\mt{\torusvar},\pi-\fb{\torusvar}\right)$.
Each of those transformations leaves the domain $\gapdom$ invariant
and exchanges between its boundary curves $\mathcal{C}_{l}$ and $\mathcal{C}_{u}$.

The direction of the flow on the torus is given by the ratio 
\begin{equation}
\slop=\frac{\mt h}{\fb h}\frac{\fb c}{\mt c}.\label{eq:flow direction}
\end{equation}
The irrationality of the ratio, on account of the physical nature
of the parameters involved, implies that the flow covers the torus
ergodically, with a uniform measure. Hence, the density of the gaps
is simply the relative area of $\gapdom$ in $\torus$, which can
be calculated via the integral of the\emph{ closed-form} expression
\begin{equation}
1-\frac{2}{\pi^{2}}\intop_{0}^{\pi}\mathcal{C}_{l}\mathrm{d}\fb{\torusvar}.\label{eq:band density}
\end{equation}
We arrive at a counter-intuitive and peculiar result; at a prescribed
$\gamma$, the gap-density is \emph{independent} of the volume fractions
of the phases. To shed light on the significance of this statement,
consider the following example. Compose a laminate of equal volume
fractions of materials $\left(1\right)$ and $\left(2\right)$. Compose
a second laminate by introducing an \emph{infinitesimal} amount of
$\left(1\right)$ into a bulk of $\left(2\right)$. The probability
that an arbitrary frequency pertains to a band-gap is \emph{identical}
in the two laminates.

The widths of the gaps are investigated next. We recognize that these
widths, which we denote by $\Delta\omega$, are related to lengths
of intervals directed along the flow, which we denote by $\Delta\zeta$,
whose endpoints lie on $\mathcal{C}_{u}$ and $\mathcal{C}_{l}$,
via
\begin{equation}
\Delta\omega=\Delta\zeta\frac{\fb{\kappa}}{\sqrt{1+\slop^{2}}},\label{eq:band gap real and on torus}
\end{equation}
where $\phase{\kappa}=\p c/\phase h$. We associate each length $\Delta\zeta$
with the parameters $a$ and $b$ which characterize the line equation
of its corresponding flow interval 
\begin{equation}
\mt{\torusvar}=a\fb{\torusvar}+b.\label{eq:torus line equation}
\end{equation}
These observations, together with the derived expressions for the
curves $\mathcal{C}_{u}$ and $\mathcal{C}_{l}$, enable determining
the width of the gaps, and relating it to the physical and geometrical
properties of the crystal. As the whole spectrum is encapsulated in
the torus, we are able, in turn, to formulate optimization problems
rigorously, derived in the sequel.

We start with the 1$^{\mathrm{st}}$ gap, whose width maximization
is of practical importance, being the one which is most often realized
experimentally \citep{Gomopoulos10nanoletters,Schneider12nano,Schneider13prl}.
We would like to know: given two materials, what is the microstructure
which maximizes the 1$^{\mathrm{st}}$ gap? The $1^{\mathrm{st}}$
gap is identified with the flow line emanating from the origin. Therefore,
we seek the slope $a$ which maximizes the right-hand side of Eq.
(\ref{eq:band gap real and on torus}), at $b=0$. The problem is
interpreted as a search for an optimal $\mt h/\fb h$ at fixed $\phase c$. 

To that end, the calculation over the torus serves as an alternative
to a calculation of the $1^{\mathrm{st}}$ gap-width for all possible
compositions, via a partial evaluation of the spectra. Calculations
over the torus become a necessity, however, when the whole spectrum
needs to be analyzed. Such cases are considered next, starting with
the calculation of the greatest gap width. We address two different
practical scenarios. The first scenario considers a given crystal,
with prescribed geometrical and physical properties. The objective
is delivered by finding the maximal $\Delta\zeta$ within the segments
along flows of the prescribed slope $a$, \emph{i.e.}, maximizing
over translations $b$, and plugging it into Eq. (\ref{eq:band gap real and on torus}).
We emphasize that this optimization is defined solely over $\gapdom$,
in an exact manner. Contrarily, a direct approach will require the
evaluation of an infinite spectrum; since in practice it must be truncated,
such approach provides only an approximation. The second scenario
we consider is of a laminate with prescribed constituents, prior to
determining their volume fractions. In this case, the maximal gap
is obtained by maximizing the right-hand side of Eq. (\ref{eq:band gap real and on torus}),
over all slopes $a$ and translations $b$. 

We supplement the discussion regarding optimality noting that our
approach allows for quick calculation of bounds on the gaps-width,
based on bounding $\Delta\zeta$. We find that for $\gamma<\gamma_{\mathrm{cr}}$,
where $\gamma_{\mathrm{cr}}\approx5.45$, the maximal $\Delta\zeta$,
denoted $\Delta\zeta_{\mathrm{max}}$, is obtained at the intersection
of $\gapdom$ with $\fb{\torusvar}=\pi/2$. Utilizing Eq. (\ref{eq:bounding curves}),
it is expressed in closed-form as 
\begin{equation}
\Delta\zeta_{\mathrm{max}}=\pi-2\arccos\left(\frac{\sqrt{\gamma^{2}-1}}{\gamma}\right),\label{eq:maximal_interval}
\end{equation}
When $\gamma>\gamma_{\mathrm{cr}}$, associated with large impedance
contrast, $\Delta\zeta_{\mathrm{max}}$ is attained along the diagonal,
for which $\fb{\torusvar}=\mt{\zeta}$. Upon substitution of $\Delta\zeta_{\mathrm{max}}$
into Eq. (\ref{eq:band gap real and on torus}), the following bound
is derived 
\begin{equation}
\Delta\omega<\frac{\fb{\kappa}\mt{\kappa}}{\sqrt{\left(\fb{\kappa}\right)^{2}+\left(\mt{\kappa}\right)^{2}}}\Delta\zeta_{\mathrm{max}}.\label{eq:bound via zeta max}
\end{equation}
Thus, an explicit expression in terms of the crystal properties is
obtained. So far, to the best of our knowledge, such bounds were not
accessible. 

The analysis is concluded with expressions for statistical characteristics
of the spectra, evaluated exactly using our representation. Specifically,
the expected value of the gap-width of a prescribed system is determined
via 
\begin{equation}
\frac{1}{\sqrt{1+a^{2}}}\fb{\kappa}\mathrm{E},\label{eq:expected value}
\end{equation}
and the variance is calculated by 
\begin{equation}
\frac{1}{\pi\left(1+a\right)\left(1+a^{2}\right)}\left(\fb{\kappa}\right)^{2}\intop_{b=-a\pi}^{b=\pi}\left[\Delta\zeta\left(b\right)-\mathrm{E}\right]^{2}\mathrm{d}b,\label{eq:variance}
\end{equation}
with $\mathrm{E}=\frac{1}{\pi\left(1+a\right)}\intop_{b=-a\pi}^{b=\pi}\Delta\zeta\left(b\right)\mathrm{d}b$.
We emphasize again that deriving these results using a direct approach
involves calculations over truncations of infinite spectra, thereby
leading to estimates rather than rigorous results.

We are ready to introduce novel composites for which our analysis
applies as well, and begin with laminates comprising non-linear soft
layers \citep{gdb05jmps,rudykh14prl}. We consider waves propagating
in such laminates when finitely deformed in a piecewise-homogeneous
manner. By applying the theory of incremental elastic deformations
\citep{ogden97book}, Eq. (\ref{eq:dispersion}) is recovered as the
dispersion relation of such ``small on large'' waves, and therefore
the resultant spectrum is endowed with the torus representation. The
physical and geometrical quantities entering Eq. (\ref{eq:dispersion})
are now functions of the finite deformation, which thus renders the
spectrum tunable. In addition to the validity of the previous results,
our representation further provides a convenient tool for characterizing
this tunability. This is demonstrated next, by way of example. The
example considers transverse waves superposed on a plane deformation
of an incompressible laminate. We denote the principle stretches in
each phase by $\phase{\lambda}$ and $1/\phase{\lambda}$, on account
of incompressibility. The lamination direction considered is along
the principal axis associated with $\phase{\lambda}$, such that $\phase h$
is related to the thickness before the deformation, $\phase H$, via
$\phase h=\phase{\lambda}\phase H$. The velocity of transverse waves
propagating in this direction is 
\begin{equation}
\phase c=\sqrt{\frac{\phase{\tilde{\mu}}}{\phase{\rho}}},\label{eq:shear velocity}
\end{equation}
where the instantaneous stiffness, $\phase{\tilde{\mu}}$, depends
on the underlying deformation and the constitutive law of the phase.
If a neo-Hookean law governs the layers behavior, we have that 
\begin{equation}
\phase{\tilde{\mu}}=\lambda^{\left(p\right)^{2}}\phase{\mu},\label{eq:mu tilde}
\end{equation}
where $\phase{\mu}$ is the shear modulus in the limit of small strains.
A simple calculation shows that $\phase{\kappa}$ and the flow direction
are unaffected by the deformation, namely, 
\begin{equation}
\phase{\kappa}=\frac{\phase c}{\phase h}=\frac{\sqrt{\lambda^{\left(p\right)^{2}}\phase{\mu}/\phase{\rho}}}{\phase{\lambda}\phase H}=\frac{\sqrt{\phase{\mu}/\phase{\rho}}}{\phase H}.\label{eq:flow direction 2}
\end{equation}
To examine how the domain of $\gapdom$ is changed by the deformation,
assume without loss of generality that $\fb{\rho}\fb{\mu}>\mt{\rho}\mt{\mu}$;
if $\mt{\fb{\lambda}>\lambda}$ (resp. $\mt{\fb{\lambda}<\lambda}$),
then $\gamma$ is greater (resp. less) than its value before the deformation,
and the area of $\gapdom$ increases (resp. decreases). Together with
the observation made on $a$ and $\phase{\kappa}$, it implies that
frequencies are added to (resp. removed from) the boundaries of the
gaps, without changing the mean frequency of each gap.

Consider next layers that are characterized by a Gentian model \citep{gent96rc&t}.
In this case, the instantaneous stiffness is 
\begin{equation}
\phase{\tilde{\mu}}=\lambda^{\left(p\right)^{2}}\phase{\mu}/\left(1-\frac{\lambda^{\left(p\right)^{2}}+\lambda^{\left(p\right)^{-2}}-2}{\p{\jm}}\right),\label{eq:mu tilde-1}
\end{equation}
where the material parameter $J_{m}\in(0,\infty]$ models the rapid
rise of stress in elastomers when approaching a limiting strain. Assume
first that $\fb{\jm}=\mt{\jm}\equiv J$ and $\fb{\lambda}=\mt{\lambda}\equiv\lambda$.
In such case, the flow direction and impedance contrast are not modified
by the deformation; however, the flow rate, $\mathrm{d}\overrightarrow{\torusvar}/\mathrm{d}\omega$,
is multiplied by $1-\left(\lambda^{^{2}}+\lambda^{^{-2}}-2\right)/J$.
Consequently, the pertinent frequencies are multiplied by the inverse
of this factor. Observing this factor is a monotonically decreasing
function of $\lambda$ with a range $(0,1]$ \footnote{The limiting strain modeled by $J_{m}$ reflects  $\lambda^{2}+\lambda^{-2}-2<J$.},
implies that the frequency spectrum is shifted towards higher frequencies,
and its gaps are rendered wider, by that factor. When $\fb{\jm}\neq\mt{\jm}$
or $\fb{\lambda}\neq\mt{\lambda}$, the flow direction changes with
the deformation, and so does the impedance contrast. Thus, the spectrum
is rendered tunable in a more intricate manner. The exploration of
the specific way it takes effect depending on the relation between
$\phase{J_{m}}$ and $\phase{\lambda}$ is beyond the scope of this
example; however, we note that following the way the aforementioned
relations enter Eq. (\ref{eq:flow direction}), the flow lines will
rotate either clockwise or counter-clockwise, and the length of their
intersections with the resultant domain of $\gapdom$ will accordingly
change. The analysis of the resultant $\Delta\omega$ is rendered
simpler, in virtue of the torus universality, as for each $\gamma$,
the relation between $\Delta\zeta$ and $a$ is unchanged. Therefore,
one can calculate the resultant slope $a$ in terms of $\phase{J_{m}}$
and $\phase{\lambda}$, and employ a fixed calculation of $\Delta\zeta/\sqrt{1+a^{2}}$,
to evaluate Eq. (\ref{eq:band gap real and on torus}). 

We continue with \emph{dielectric elastomers} (DEs), a class of soft
active materials. DEs undergo finite strains and change their physical
properties by application of electric stimuli \citep{pelr&etal00scie,suo10prl}.
The work in Ref. \citep{gg12} has shown that under particular settings,
the propagation of waves in finitely deformed DE laminates is also
described by Eq. (\ref{eq:dispersion}). Hence, the torus representation
holds in this case as well, and subsequently, the validity of our
previous observations is established. In particular, the torus representation
provides a platform for an investigation of the effect the electric
field has on the frequency spectrum. 

We consider next \emph{magnetorehological elastomers }(MREs) \citep{ginderetal99MRE}\emph{.
}These materials consist of ferromagnetic particles embedded in a
rubber-like matrix. Application of magnetic stimuli induces magnetic
forces and moments on the inclusions. This changes the microstructure
of the material, and, in turn, alters its configuration and stiffness.
When the particles are disturbed in a chain-like manner, the material
admits a laminated structure \citep{Galipeau08102013,Rudykh2013jmps}.
We argue that our analysis also applies to MREs, in light of the similarly
with DEs. First, we note that under a quasi magneto/electrostatic
approximation, the governing electric and magnetic fields are differentially
similar \footnote{The magnetic induction and the electric displacement fields are divergence-free,
the magnetic and the electric fields are curl-free.}. Second, the mechanical response of both MREs and DEs is governed
by an elastomeric substance. The only difference is the relation between
the magnetic load and the resultant stretch, on account of a different
magnetic constitutive behavior. Therefore, under the same settings,
the propagation of superposed waves is governed by the same dispersion
relation derived for DEs. The applicability of our analysis thereby
follows.

We complete this letter with a summary of our main conclusions, and
a glance towards future challenges. We found that the frequency spectra
of various 1D crystals admit a universal structure, independent of
the geometry of their unit-cells and specific physical properties.
This structure enabled us to derive universal properties of the spectra,
and rigorously determine the gaps-density, their expected and maximal
widths, and relate these to particular compositions. We showed that
our conclusions successfully apply to certain multi-physical materials,
of tunable spectra. We utilized our framework to characterize this
tunability. Thus far, the results above were either unknown, or determined
in an approximate manner. 

Our future objectives are to extend this approach to analyze interesting
generalizations of the considered systems. One such generalization
is a crystal consisting of more than two constituents. In this case,
the dispersion relation will depend on additional products \citep{shen00}
in the form of Eq. (\ref{eq:function for torus}), as inferred from
a transfer-matrix analysis \citep{shen00,gg12}. Our extended analysis
will consequently require a torus whose dimensionality equals the
number of constituents. $\mathbb{N}$-periodic media are also systems
of major interest. In these composites, a closed-form expression for
the dispersion relation is not available, and series-type solutions
are sought \citep{Kushwaha1993,Vasseur08}. It is an imperative challenge
to establish counterparts of the results reported herein for such
materials. 

G.S. acknowledges the support of the Israel Science Foundation, founded
by the Israel Academy of Sciences and Humanities (Grant 1912/15),
and the United States-Israel Binational Science Foundation (Grant
No. 2014358). R.B. was supported by ISF (grant No. 494/14), Marie
Curie Actions (grant No. PCIG13-GA-2013-618468) and the Taub Foundations
(Taub Fellow).

\bibliographystyle{unsrtnat}

\end{document}